\documentclass[aps,amsfonts,superscriptaddress,showpacs,floatfix]{revtex4}
\usepackage{amsmath,amsthm,amssymb}
\usepackage{graphicx}
\usepackage{dcolumn}
\usepackage{bm}
\usepackage{natbib}
\begin{document}
\newtheorem{thm}{Theorem}[section]
\newtheorem{cor}[thm]{Corollary}
\newtheorem{lem}[thm]{Lemma}
\title{Random matrix averages and the impenetrable Bose gas in
  Dirichlet and Neumann boundary conditions}

\author{P.J. Forrester}
\email[]{P.Forrester@ms.unimelb.edu.au}
\affiliation{Department of Mathematics and Statistics, University of Melbourne,
Victoria 3010, Australia}
\author{N.E. Frankel}
\email[]{n.frankel@physics.unimelb.edu.au}
\affiliation{School of Physics, University of Melbourne,
Victoria 3010, Australia}
\author{T.M. Garoni}
\email[]{t.garoni@physics.unimelb.edu.au}
\affiliation{School of Physics, University of Melbourne, 
Victoria 3010, Australia}

\date{\today}
\newcommand{\norm}{\bar\lambda_0}
\newcommand{\sgn}{\text{\text{\text sgn}}}
\newcommand{\ds}{\displaystyle}
\newcommand{\ts}{\textstyle}
\newcommand{\be}{\begin{equation}}
\newcommand{\ee}{\end{equation}}
\newcommand{\ba}{\begin{eqnarray}}
\newcommand{\ea}{\end{eqnarray}}
\newcommand{\bi}{\bibitem}
\newcommand{\intl}{\int\displaylimits}
\newcommand{\suml}{\sum\displaylimits}
\newcommand{\prodl}{\prod\limits}
\newcommand{\cupl}{\cup\displaylimits}
\newcommand{\bl}{\left(}
\newcommand{\br}{\right)}
\newcommand{\wf}{\psi_N^C(x_1,x_2,...,x_N)}
\newcommand{\dm}{\varrho^C_{N}(t)}
\newcommand{\mom}{c_n(N)}
\newcommand{\dmf}{\varrho_N^{C,F}(t)}
\newcommand{\wdm}{\rho^H_N(x,y)}
\newcommand{\elem}{b_{j,k}(x,y)}
\newcommand{\w}{\sqrt{2N}}
\newcommand{\s}{S_n(\lambda_1,\lambda_2,\lambda)}
\newcommand{\pdfs}{p.d.f.'s }
\newcommand{\C}{C_j^{1/4}}
\newcommand{\pdf}{p.d.f. }
\newcommand{\ddm}{\rho_{N+1}^D(x,y)}
\begin{abstract}
The density matrix for the impenetrable Bose gas in Dirichlet
and Neumann boundary conditions can be written in terms of
$\left<\prod_{l=1}^n\vert \cos\phi_1-\cos\theta_l\vert
  \vert\cos\phi_2-\cos\theta_l\vert\right>$, where the average is with
respect to the eigenvalue probability density function for random
unitary matrices from the classical groups $Sp(n)$ and $O^+(2n)$
respectively. In the large $n$ limit log-gas considerations imply that
the average factorizes into the product of averages of the form
$\left<\prod_{l=1}^n\vert\cos\phi-\cos\theta_l\vert\right>$.
By changing variables this average in turn is a special case of the
function of $t$ obtained by averaging $\prod_{l=1}^n\vert
t-x_l\vert^{2q}$ over the Jacobi unitary ensemble from random matrix
theory. The latter task is accomplished by a duality formula from the
theory of Selberg correlation integrals, and the large $n$ asymptotic
form is obtained. The corresponding large $n$ asymptotic form of the
density matrix is used, via the exact solution of a particular
integral equation, to compute the asymptotic form of the low lying
effective single particle states and their occupations, which are
proportional to $\sqrt{N}$.
\end{abstract}


\maketitle

\section{Introduction}
\label{introduction}
The probability density functions (\pdfs) 
\be
{1\over n!}\left({1\over 2\pi}\right)^n \prod_{l=1}^n 4 \sin^2(\theta_l)
\prod_{1\le j<k\le n}
4 (\cos\theta_k -\cos\theta_j)^2
\label{symplectic}
\ee
\be
{2\over n!}\left({1\over 2\pi}\right)^n 
\prod_{1\le j<k\le n}
4 (\cos\theta_k -\cos\theta_j)^2
\label{orthogonal}
\ee
where $0\le \theta_j\le \pi$ $(j=1,...n)$ occur in both random matrix
theory and the quantum many body problem. In the former
they are eigenvalue \pdfs for classical groups with the Haar (uniform)
measure \--- the group $Sp(n)$ of $n\times n$ unitary matrices with real
quaternion elements (which are themselves $2\times 2$ matrices), and
the group $O^+(2n)$ of $2n\times 2n$ unitary matrices with real
elements (real orthogonal matrices) and determinant equal to $+1$, for
(\ref{symplectic}) and (\ref{orthogonal}) respectively. A self
contained derivation of these facts can be found in \cite[Chapter
2]{ForresterBook}. In the latter
they are the absolute value squared of the ground state wave function for $n$ free
fermions on the interval $[0,\pi]$ with Dirichlet and Neumann
boundary conditions respectively. As is similarly well known, and
revised from first principles in our work \cite{cmp}, they are also the absolute value
squared of the ground state wave function for $n$ impenetrable bosons on the
interval $[0,\pi]$ \--- in one-dimension the ground state wave function of the
impenetrable Bose system is equal to the absolute value of the
corresponding free Fermi system.

In studies relating to both these seemingly disparate interpretations
of the \pdfs (\ref{symplectic}) and (\ref{orthogonal}) there is cause
to investigate the function of $(\phi,m,n)$ defined by averaging
\be
\prod_{l=1}^n
(\cos\phi -\cos\theta_l)^m
\label{Lfnaverageand}
\ee 
with respect to these \pdfs. In the random matrix interpretation this
comes about in applications to $L$-function theory
\cite{KatzSarnakBook,KeatingSnaithPaper,KeatingSnaithReview03}.
Briefly, there are families of $L$-functions with special symmetries
which are known to have their non-trivial zeros well described by
eigenvalues of random matrices from the classical group corresponding
to that symmetry. For $\cos\phi=1$ and small values of $m$ the
expected value of (\ref{Lfnaverageand}) can be computed with
$\{\theta_l\}$ corresponding to the zeros of particular families of
$L$-functions, and it can also be computed \--- with $m$ a general
non-negative integer \--- for the random matrix ensembles. This then
allows for both a test of the original hypothesis relating
$L$-functions to random matrices, and provides specific conjectures
for the statistical properties of the zeros of the $L$-function
families.

In the quantum many body interpretation the immediate interest is not
in the average of (\ref{Lfnaverageand}), but rather the average of 
\be
\prod_{l=1}^n
\vert\cos\phi_1 -\cos\theta_l\vert\vert\cos\phi_2 -\cos\theta_l\vert.
\label{qmaverageand}
\ee 
This gives the ground state density matrix of the corresponding
impenetrable Bose gas system (if the absolute value signs are removed,
the average gives the ground state density matrix for the free Fermi
system) \cite{cmp}. However the study of (\ref{qmaverageand}) leads
back to the computation of (\ref{Lfnaverageand}). Thus as noted in
\cite{Forrester92} and \cite{pra} for the problem of computing the
asymptotic behavior of the ground state density matrix for the
impenetrable bosons in the bulk and in an harmonic trap respectively,
for large $n$ and $\phi_1$ and $\phi_2$ fixed the average
(\ref{qmaverageand}) is expected to factorize, and be proportional to
the average of the product involving $\phi_1$ times the average of
the product involving $\phi_2$. The latter are then the continuation
in $m$ of (\ref{Lfnaverageand}) from the even positive integers to the
value $m=1$ (this being  a way to effectively study the average of the
absolute value of (\ref{Lfnaverageand})).

We remark that in the case of the density matrix for the impenetrable
bosons in periodic boundary conditions the analogous task is to
compute the average of
\be
\prod_{l=1}^n\vert e^{i\phi_1}-e^{i\theta_l}\vert \,\vert e^{i\phi_2}-e^{i\theta_l}\vert
\ee
with respect to the p.d.f.
\be
\frac{1}{n!}\left(\frac{1}{2\pi}\right)^n
\prod_{1\le j<k\le n}\vert e^{i\theta_k}-e^{i\theta_j}\vert^2,
\ee
where $-\pi<\theta_j\le \pi$ $(j=1,...,n)$. In this case the
asymptotic form for large $n$ with $\phi_1$ and $\phi_2$ fixed is a
special case of known asymptotic forms for Toeplitz determinants with
singular generating functions of the so called Fisher-Hartwig type
(for an extended discussion on this point and references to the
relevant literature see \cite{cmp}). The average of (\ref{qmaverageand})
with respect to (\ref{symplectic}) or (\ref{orthogonal}) can readily be written as a
Hankel determinant, but there is no known analogue of the
Fisher-Hartwig asymptotic form.

In Section \ref{Duality formulas} of this work we show how, for $m$
an even positive integer, the average of (\ref{Lfnaverageand}) with
respect to the \pdfs (\ref{symplectic}) and (\ref{orthogonal}), which
is by definition an $n$-dimensional integral, can be written as a
$m$-dimensional integral. From the latter the large $n$ asymptotic
form of the average is deduced. In Section \ref{Asymptotic form} the
result of Section \ref{Duality formulas} is used to deduce the large
$n$ asymptotic form of the average (\ref{qmaverageand}) with respect
to the \pdfs (\ref{symplectic}) and (\ref{orthogonal}) and thus of the
ground state density matrix. We conclude in Section \ref{implications} by applying
our result for the asymptotic form of the density matrix to the
computation of the occupations of the effective single particle
states. We also make some remarks in relation to the wider setting of
our asymptotic analysis, in which we use Coulomb gas arguments to
formulate an analogue of the Fisher-Hartwig asymptotic form for a
class of Jacobi unitary ensemble averages. 

\section{Duality formulas for multiple integrals and asymptotic
  analysis}
\label{Duality formulas}
\subsection{The duality formula}
The change of variables 
\be
x_j={1\over 2}(\cos\theta_j +1)\qquad (0\le x_j \le 1, \quad j=1,...n)
\label{change}
\ee
transforms (\ref{symplectic}) and (\ref{orthogonal}) into the \pdfs 
\be
{1\over n!}\left({1\over 2\pi}\right)^n \prod_{l=1}^n (x_l(1-x_l))^{1/2}
\prod_{1\le j<k\le n} (x_k -x_j)^2
\label{dirichlet}
\ee
\be
{2\over n!}\left({1\over 2\pi}\right)^n \prod_{l=1}^n (x_l(1-x_l))^{-1/2}
\prod_{1\le j<k\le n} (x_k -x_j)^2
\label{neumann}
\ee
respectively. These \pdfs in turn are special cases of the class of
\pdfs proportional to 
\be
\prod_{l=1}^n
x_l^{\lambda_1}(1-x_l)^{\lambda_2}\prod_{1\le j<k\le n}(x_k-x_j)^2
\label{jue}
\ee
known in random matrix theory as the Jacobi unitary ensemble. Also,
under the change of variables (\ref{change}) the task of computing the
average of (\ref{Lfnaverageand}) becomes the task of computing the
average of 
\be
\prod_{l=1}^n(t-x_l)^m
\label{charfn}
\ee
with respect to (\ref{dirichlet}) and (\ref{neumann}), or more
generally with respect to (\ref{jue}).

In fact there is an advantage in further generalizing the setting of
the computation of the average of  (\ref{charfn}) and considering the
class of multiple integrals known as Selberg correlation integrals,
defined by
\begin{align}
S_{n,m}(\lambda_1,\lambda_2,\lambda;t_1,...,t_m)
& :=
{1\over C}\int_{[0,1]^n}dx_1...dx_n\prod_{l=1}^n
\left(x_l^{\lambda_1}(1-x_l)^{\lambda_2}\prod_{i=1}^m(t_i-x_l)\right)
\nonumber \\
& \phantom{:=} \,\times 
\prod_{1\le j<k\le n}\vert x_k-x_j\vert^{2\lambda}
\nonumber \\
& =
\left<\prod_{l=1}^n\prod_{i=1}^m(t_i-x_l)\right>_{\text{J}(2\lambda)\text{E}_n}.
\label{SelbergCorrelation}
\end{align}
Here 
\be
C=\s=
\int_{[0,1]^n}dx_1...dx_n
\prod_{l=1}^nx_l^{\lambda_1}(1-x_l)^{\lambda_2}\prod_{1\le j < k\le n}
\vert x_k -x_j\vert^{2\lambda},
\label{selberg}
\ee
known as the Selberg integral, is the normalization chosen so that the
coefficient of $\prod_{i=1}^mt_i^n$ is unity, and the average over
$\textnormal{J}(2\lambda)\textnormal{E}_n$ refers to the \pdf
\be
{1\over\s}\prod_{l=1}^nx_l^{\lambda_1}(1-x_l)^{\lambda_2}\prod_{1\le j < k\le n}
\vert x_k -x_j\vert^{2\lambda}
\label{generalisedJUE}
\ee
(the notation $\textnormal{J}(2\lambda)\textnormal{E}_n$ denotes the Jacobi -$(2\lambda)$
ensemble, which with $\lambda=1$ corresponds to the Jacobi unitary
ensemble (\ref{jue})). Setting $t_1=...=t_m=t$ in
(\ref{SelbergCorrelation}) gives the average of (\ref{charfn}) with
respect to (\ref{generalisedJUE}). The advantage in studying
(\ref{SelbergCorrelation}) is that we can put to use the discovery of
\cite{Kaneko93}, relating the Selberg correlation integrals to the
theory of Jack polynomials (in the case $\lambda=1$ the Jack polynomials
coincide with the Schur polynomials \cite{Macdonald95}). In particular
the Selberg correlation integrals were evaluated in terms of a
generalization of the Gauss hypergeometric function $_2F_1$ based on
the Jack polynomials. It was realized by one of the present authors
\cite{Forrester93a,Forrester93b,Forrester94,BakerForrester97,Forrester98,ForresterWitte02} that
theory initiated in \cite{Kaneko93} could be further developed and
used to express the average of (\ref{neumann}) with respect to
(\ref{generalisedJUE}) and its limiting forms as the Laguerre -
$(2\lambda)$ ensemble and the Gaussian - $(2\lambda)$ ensemble, as $m$
- dimensional integrals. Because the role of $n$ and $m$ is
effectively interchanged, these integration identities have been
referred to as duality formulas
\cite{ForresterWitte01,BrezinHikami01,MulaseWaldron02}. One of their uses,
as we will demonstrate in the case of the average of (\ref{charfn})
with respect to (\ref{jue}), is in the computation of the large $n$ asymptotics.

The particular duality formula of interest to us is given explicitly
in \cite{ForresterWitte02}. To state the result we must
introduce the generalized circular ensemble, $\textnormal{C}\beta \textnormal{E}_N$, as the
\pdf proportional to 
\be
\prod_{1\le j<k\le N} \vert z_k -z_j\vert^{\beta} ,\qquad
(z_j=e^{i\theta_j},\,\,-\pi<\theta_j<\pi, \,\,j=1,...,N)
\label{generalisedCE}
\ee
With this notation, we read off from \cite{ForresterWitte02}
eq. (3.41) that
 \be
 \left<\prod_{l=1}^N z_l^{(\eta_1-\eta_2)/2}
 \vert 1+z_l\vert^{\eta_1+\eta_2}
 (1+t z_l)^m
 \right>_{\textnormal{C}\beta \textnormal{E}_N}
 \propto
 \left<
 \prod_{l=1}^m[1-(1-t)x_l]^N
 \right>_{\textnormal{J}(4/\beta)\textnormal{E}_m}
 {\Bigg\vert}_{\ts{\lambda_1=2(\eta_2-m+1)/\beta -1\atop \lambda_2=2(\eta_1+1)/\beta-1}}
\label{circular to jacobi} 
\ee
But we want to make
$\left<\prod_{l=1}^n(t-x_l)^m\right>_{\text{J}(2\lambda)\text{E}_n}$ the quantity
being transformed, so (\ref{circular to jacobi}) requires manipulation. For this we write
\be
t\mapsto 1-{1\over t}, \qquad m\leftrightarrow N, \qquad {2\over
  \beta}=\lambda,\qquad N=n
\ee
Noting that then
\be
\eta_1={1\over \lambda}(\lambda_2+1)-1,\qquad\eta_2={1\over\lambda}(\lambda_1+1)+n-1,
\label{etadefinitions}
\ee
multiplying both sides of (\ref{circular to jacobi}) by $t^{m N}$ and taking the complex
conjugate of the left hand side of (\ref{circular to jacobi}) shows 
\be
\left<\prod_{l=1}^n(t-x_l)^m
\right>_{\textnormal{J}(2\lambda)\textnormal{E}_n}
=
A\left<\prod_{l=1}^mz_l^{([(\lambda_1-\lambda_2)/\lambda]-n)/2}\vert
  1+z_l\vert^{[(\lambda_1+\lambda_2+2)/\lambda]+n -2}[t(1+z_j)-1]^n
\right>_{\text{C}(2/\lambda)\text{E}_m}
\label{jacobitocircular}
\ee
where A, the proportionality constant, is independent of $t$. To
specify A requires, in addition to the Selberg integral
(\ref{selberg}), the so called Morris integral 
\be
M_n(a,b,\lambda):=\left({1\over 2\pi}\right)^n
\int_{-\pi}^{\pi}d\theta_1...\int_{-\pi}^{\pi}d\theta_n
\prod_{l=1}^n
z_l^{(a-b)/2}\vert 1+z_l\vert^{a+b}
\prod_{1\le j<k\le n}
\vert z_k-z_j\vert^{2\lambda}.
\ee
Then setting $t=1$ in (\ref{jacobitocircular}) shows that 
\be
A={S_n(\lambda_1,\lambda_2+m,\lambda)\over S_n(\lambda_1,\lambda_2,\lambda)}
{M_m(0,0,1/\lambda)\over M_m(\eta_2,\eta_1,1/\lambda)}
\label{A}
\ee
where $\eta_1$ and $\eta_2$ are given by (\ref{etadefinitions}). Both
the Selberg integral and Morris integral have exact evaluations in
terms of products of gamma functions (see e.g. \cite{ForresterBook}). In
the case $\lambda=1$ these read
\ba
S_n(a,b,1)
&=&
\prod_{j=0}^{n-1}
{\Gamma(a+1+j)\Gamma(b+1+j)\Gamma(2+j)\over\Gamma(a+b+1+n+j)}
\nonumber \\
&=&
{G(n+1+a)\over G(1+a)}
{G(n+1+b)\over G(1+b)}
{G(n+1+a+b)\over G(2n+1+a+b)}G(n+2),
\label{selberg of 1}
\\
M_n(a,b,1)
&=&
\prod_{j=0}^{n-1}
{\Gamma(a+b+1+j)\Gamma(2+j)\over \Gamma(a+1+j)\Gamma(b+1+j)}
\nonumber \\
&=&
{G(n+1+a+b)\over G(1+a+b)}
{G(1+a)\over G(n+1+a)}
{G(1+b)\over G(n+1+b)}G(n+2),
\label{morris of 1}
\ea
where $G(z)$ denotes the Barnes $G$-function, related to the gamma function
by the functional equation 
\be
G(z+1)=\Gamma(z)G(z).
\label{BarnesToGamma}
\ee
\subsection{Asymptotics}
Our interest is in the asymptotic form of the $\text{J}(2\lambda)\text{E}_n$ average
in (\ref{jacobitocircular}) in the case $\lambda=1$ and $m$ even. The
experience of our previous study \cite{pra}, in which we studied the
same product averaged over the Gaussian unitary ensemble (eigenvalue
\pdf of complex Hermitian matrices with Gaussian entries) using a
result known in the literature \cite{BrezinHikami00} (see also \cite{StrahovFyodorov02}), tells us the
related quantity
\be
{Z_{n,\lambda_1,\lambda_2}((X,q))\over Z_{n+q,\lambda_1,\lambda_2}((\cdot,0))}
\label{partitionratio}
\ee
where 
\ba
Z_{n,\lambda_1,\lambda_2}((X,q))
&=&
X^{\lambda_1 q}(1-X)^{\lambda_2 q}
\int_0^1dX_1...\int_0^1dX_n
\prod_{l=1}^n X_l^{\lambda_1}(1-X_l)^{\lambda_2}
\vert X-X_l\vert^{2q}
\nonumber \\
&\times &
\prod_{1\le j<k \le n}\vert X_k-X_j\vert^2
\label{partition definition}
\ea
is better suited for the purpose. In addition to  multiplying the
$\textnormal{J}\textnormal{U}\textnormal{E}_n$ average in (\ref{jacobitocircular}) by the $t$-dependent
factor $t^{\lambda_1 m/2}(1-t)^{\lambda_2 m/2} $, a key feature of 
(\ref{partitionratio}) is the normalization chosen so that in the
interpretation of (\ref{partition definition}) as the configuration
integral for a log-potential Coulomb gas the  normalization has the
same total charge.

It follows from (\ref{jacobitocircular}), (\ref{A}), and noting 
\be
Z_{n+q,\lambda_1,\lambda_2}((\cdot,0))=S_{n+q}(\lambda_1,\lambda_2,1),
\ee
that we have the duality formula for (\ref{partitionratio}) with
$q=m/2$, $m$ even 

\begin{align}
{Z_{n,\lambda_1,\lambda_2}((t,m/2))\over Z_{n+m/2,\lambda_1,\lambda_2}((\cdot,0))}
&=
{S_n(\lambda_1,\lambda_2+m,1)
\over
S_{n+m/2}(\lambda_1,\lambda_2,1) M_m(\lambda_1+n,\lambda_2,1)}
t^{\lambda_1 m/2}(1-t)^{\lambda_2 m/2} 
\nonumber \\
& \phantom{=} \,\times
\int_{-\pi}^{\pi}d\theta_1...\int_{-\pi}^{\pi} d\theta_m
\prod_{l=1}^m
z_l^{(\lambda_1-\lambda_2-n)/2}
\vert 1+z_l\vert^{\lambda_1+\lambda_2+n}
[t(1+z_l)-1]^n
\nonumber \\
& \phantom{=} \,\times
\prod_{1\le j<k\le m}
\vert z_k -z_j\vert^2
\nonumber \\
&=
{S_n(\lambda_1,\lambda_2+m,1)\over (2\pi i)^m
  S_{n+m/2}(\lambda_1,\lambda_2,1)M_m(\lambda_1+n,\lambda_2,1)}
I_n(t),
\label{partition ratio to I}
\end{align}
where 
\ba
I_n(t)&:=&
t^{\lambda_1 m/2}(1-t)^{\lambda_2 m/2}
\int_{{\cal{C}}^m}dx_1...dx_m\prod_{l=1}^m(-x_l)^{\lambda_1+\lambda_2+n}
(1-x_l)^{-(\lambda_2+n+m)}
(1-t x_l)^n
\nonumber \\
&\times &
\prod_{1\le j<k\le m}(x_k-x_j)^2.
\label{I}
\ea
In (\ref{I}) ${\cal C}$ is any simple closed contour starting at
$x_j=0$ in the complex plane and encircling $x_j=1$ anti-clockwise
without crossing the interval $x_j\in (0,1)$. We obtain the second
equality in (\ref{partition ratio to I}) by writing the integrand in a
form without absolute value signs, changing variables
\be
d\theta_j = {1\over 2\pi iz_j}dz_j
\ee 
then noticing the integrand is analytic except at $z_j=0,1$ and with a
cut along the interval $z_j\in(0,1)$. 

The large $n$, fixed $t\in(0,1)$ asymptotic analysis of an integral
very similar to (\ref{I}) has been detailed in \cite{Forrester98}, and
that analysis in turn follows the stationary phase analysis of a
related multiple integral given in \cite{Forrester93b}. Now, the
$n$-dependent terms in the integrand of $I_n(t)$ are 
\be
x_j^n(1-x_j)^{-n}(1-tx_j)^n=\exp[n(\log x_j-\log(1-x_j)+\log(1-tx_j))].
\label{saddle}
\ee
As noted in \cite{Forrester98}, a simple calculation shows that the
stationary point of the exponent occurs when 
\be
x_j=1\pm i\left[{1\over t}(1-t)\right]^{1/2}=:x_{\pm}.
\ee
This suggests we deform the contours so that $m/2$ integration
variables, $(x_1,...,x_{m/2})$ say, pass through $x_+$, and the
remaining pass through $x_-$. We must then expand the integrand in the
neighborhood of these stationary points. Because we have made a definite choice
of the $m/2$ variables, we must multiply by the combinatorial factor
$\left({m\atop m/2}\right)$. From \cite{Forrester98} we know that
expanding the exponent in (\ref{saddle}) to second order in
$(x_{\pm}-x_j)$ gives
\be
\exp[n(\log x_j-\log(1-x_j) +\log(1-t x_j))]\sim
\exp[-{n\alpha\over 2}(x_j-x_{\pm})^2]
\ee
where
\be
\alpha={t^2\over (1-tx_{\pm})^2}+{1\over x_{\pm}^2}-{1\over
(1-x_{\pm})^2}.
\ee
Regarding the leading order expansion of the other terms in the
integrand we have 
\be
\prod_{1\le j<k\le m}(x_k-x_j)^2\sim (x_+-x_-)^{2(m/2)^2}\prod_{1\le
j<k\le m/2}
(x_k-x_j)^2\prod_{m/2+1\le j<k\le m}(x_k-x_j)^2,
\ee
\be
\prod_{j=1}^m(-x_j)^{\lambda_1+\lambda_2}(1-x_j)^{-(\lambda_2
+m)}\sim
\vert x_+\vert^{(\lambda_1 +\lambda_2)m}\vert 1-x_+\vert^{-(\lambda_2+m)m}.
\ee
Hence
\begin{align}
I_n(t)
&\sim \left({m\atop m/2}\right)
[t^{\lambda_1}(1-t)^{\lambda_2}]^{m/2}
(x_+-x_-)^{2(m/2)^2}\vert x_+\vert^{(\lambda_1+\lambda_2)m}
\vert 1-x_+\vert^{-(\lambda_2+m)m}
\nonumber \\
& \phantom{=} \,\times
{\Bigg\vert}
\int_{-\infty}^{\infty}dx_1...\int_{-\infty}^{\infty}dx_{m/2}
\prod_{l=1}^{m/2}
\exp\left[{-n\alpha\over 2}x_l^2\right]
\prod_{1\le j<k\le {m/2}}
(x_k-x_j)^2
{\Bigg\vert}^2
\nonumber \\
&=
\left({m\atop m/2}\right)[t^{\lambda_1}(1-t)^{\lambda_2}]^{m/2}
(x_+-x_-)^{2(m/2)^2}\vert x_+\vert^{(\lambda_1+\lambda_2)m}
\vert 1-x_+\vert^{-(\lambda_2+m)m}
\nonumber \\
& \phantom{=} \,\times
\vert n\alpha\vert^{-(m/2)^2}(V_{m/2})^2
\label{ItoV}
\end{align}
where
\be
V_{m/2}:=\int_{-\infty}^{\infty}dx_1...\int_{-\infty}^{\infty}dx_{m/2}\prod_{l=1}^{m/2}
\exp\left[-{1\over 2}x_l^2\right]
\prod_{1\le j<k\le {m/2}}
(x_k-x_j)^2.
\ee

From \cite{Forrester98} 
\be
\vert x_+\vert =\sqrt{1\over t}, \quad  \vert 1-x_+\vert=\sqrt{1-t\over
t},\quad
\vert x_+-x_-\vert=2\sqrt{1-t\over t}, 
\quad
\vert\alpha\vert={2t^{3/2}\over (1-t)^{1/2}}
\ee
so (\ref{ItoV}) simplifies to read 
\be
I_n(t)\sim (-1)^{m/2}\left({m\atop m/2}\right)2^{(m/2)^2} n^{-(m/2)^2}
[t(1-t)]^{-m^2/8}(V_{m/2})^2.
\ee
Furthermore, we recognize $V_{m/2}$ as a limiting case of the Selberg
integral known as the Mehta integral \cite{Mehta91}, which has the
evaluation 
\be
V_{m/2}=(2\pi)^{m/4}\prod_{j=0}^{m/2-1}\Gamma(2 +j)=
(2\pi)^{m/4}G(m/2+2).
\label{VtoG}
\ee

Recalling (\ref{partition ratio to I}), our remaining task is to compute
the asymptotic form of the combination of Selberg integrals and the
Morris integral therein. According to (\ref{selberg of 1}) and
(\ref{morris of 1}), for this we require knowledge of the asymptotic
expansion of the Barnes $G$-function. In fact Barnes himself showed
\cite{Barnes00}
\be
\log\left({G(n+a+1)\over G(n+b+1)}\right)
\operatornamewithlimits{\sim}_{n\to\infty}(b-a)n+{a-b\over 2}\log(2\pi) + \left((a-b)n
+{a^2-b^2\over 2}\right)\log n +o(1).
\ee
Using this we find 
\be
{S_n(\lambda_1,\lambda_2+m,1)\over
S_{n+m/2}(\lambda_1,\lambda_2,1)
M_m(\lambda_1+n,\lambda_2,1)}
\sim {n^{m^2/2-m/2}\over G(m+2)}.
\label{Selberg ratio to G}
\ee
Substituting (\ref{VtoG}) in (\ref{ItoV}), then substituting the
result together with  (\ref{Selberg ratio to G}) in
(\ref{partition ratio to I}) we obtain the sought asymptotic formula
\be
{Z_{n,\lambda_1,\lambda_2}((t,q))\over
Z_{n+q,\lambda_1,\lambda_2}((\cdot,0))}
\sim
{1\over \pi^q}{G^2(q+1)\over G(2q+1)}(2n)^{-q+q^2}[t(1-t)]^{-q^2/2},
\label{asymptotic partition ratio}
\ee
where we have set $m/2=q$ and use has been made of the functional
equation (\ref{BarnesToGamma}). We note that the right hand side of
(\ref{asymptotic partition ratio}) is independent of the parameters
$\lambda_1$ and $\lambda_2$.

A check on our workings to this stage is the special case $q=1$. Then
(\ref{partitionratio})
coincides with the eigenvalue density in the JUE, $\rho^{\textnormal{J}\textnormal{U}\textnormal{E}}(X)$,
normalized so that its integral on $[0,1]$ is unity. Setting $q=1$ in
(\ref{asymptotic partition ratio}) we read off that
\be
\rho^{\textnormal{J}\textnormal{U}\textnormal{E}}(X)\sim {1\over\pi}(X(1-X))^{-1/2},
\ee
which is indeed the known functional form (see e.g. \cite{ForresterBook}). 
\section{Asymptotic form of the density matrices}
\label{Asymptotic form}
Consider the impenetrable Bose gas of $N+1$ particles confined to the
interval $[0,L]$ with Dirichlet boundary conditions. We know from
\cite{cmp} that the ground state density matrix $\ddm$ is
given by
\ba
\ddm 
&=&
{(N+1)\over C}\left(\sin{\pi x\over L}\right)\left(\sin{\pi y\over L}\right)
\int_0^Ldx_1...\int_0^Ldx_N
\prod_{l=1}^N\sin^2{\pi x_l\over L}
\nonumber \\
&\times&
{\Big\vert} \cos{\pi x\over L}-\cos{\pi x_l\over L}{\Big\vert}
{\Big\vert} \cos{\pi y\over L}-\cos{\pi x_l\over L}{\Big\vert}
\prod_{1\le j<k\le N}
{\Big\vert}\cos{\pi x_k\over L}-\cos{\pi x_j\over L}{\Big\vert}^2
\label{ddm}
\ea
where
\be
C=\int_0^Ldx_1...\int_0^Ldx_{N+1}
\prod_{l=1}^{N+1}\sin^2{\pi x_l\over L}\prod_{1\le j<k\le N+1}
{\Big\vert}\cos{\pi x_k\over L}-\cos{\pi x_j\over L}{\Big\vert}^2.
\label{dC}
\ee
Let us now change variables 
\be
\cos{\pi x_j\over L}=2X_j -1
\label{change variables}
\ee
in both (\ref{ddm}) and (\ref{dC}), and let us define
\be
\rho_{N+1}^D(X,Y):=\rho_{N+1}^D(x,y){\Big\vert}_{\ts{\cos\pi x/L=2X-1\atop \cos\pi y/L=2Y-1}}.  
\label{D scaled density matrix}
\ee
Then in terms of the generalization of (\ref{partition definition}), 
\ba
Z_{n,\lambda_1,\lambda_2}((X,q_1),(Y,q_2))&=&
\vert
X-Y\vert^{2q_1q_2}X^{\lambda_1q_1}
(1-X)^{\lambda_2q_1}Y^{\lambda_1q_2}
(1-Y)^{\lambda_2q_2}\int_0^1dX_1...\int_0^1dX_n
\nonumber \\
&\times&
\prod_{l=1}^n
X_l^{\lambda_1}(1-X_l)^{\lambda_2}\vert X-X_l\vert^{2q_1}\vert Y-X_l\vert^{2q_2}
\prod_{1\le j<k\le n}
\vert X_k-X_j\vert^2
\label{generalized partition}
\ea
we have
\be
\rho_{N+1}^D(X,Y)
= {\pi\rho\over\vert X -Y\vert^{1/2}}[X(1-X)]^{1/4}[Y(1-Y)]^{1/4}
{Z_{N,1/2,1/2}((X,1/2),(Y,1/2))\over Z_{N+1,1/2,1/2}((\cdot,0),(\cdot,0))},
\label{D partition ratio}
\ee
where $\rho:=N/L$.

Similar considerations apply to the impenetrable Bose gas of $N+1$
particles confined to the interval $[0,L]$ with Neumann boundary
conditions. Again from \cite{cmp} we know that the ground state
density matrix $\rho_{N+1}^N(x,y)$ is given by
\ba
\rho^N_{N+1}(x,y) 
&=&
{(N+1)\over C}
\int_0^Ldx_1...\int_0^Ldx_N
\prod_{l=1}^N
{\Big\vert} \cos{\pi x\over L}-\cos{\pi x_l\over L}{\Big\vert}
{\Big\vert} \cos{\pi y\over L}-\cos{\pi x_l\over L}{\Big\vert}
\nonumber \\
&\times&
\prod_{1\le j<k\le N}
{\Big\vert}\cos{\pi x_k\over L}-\cos{\pi x_j\over L}{\Big\vert}^2
\label{ndm}
\ea
where
\be
C=\int_0^Ldx_1...\int_0^Ldx_{N+1}
\prod_{1\le j<k\le N+1}
{\Big\vert}\cos{\pi x_k\over L}-\cos{\pi x_j\over L}{\Big\vert}^2.
\label{nC}
\ee
Defining
\be
\rho_{N+1}^N(X,Y):=\rho_{N+1}^N(x,y){\Big\vert}_{\ts{ \cos\pi x/L=2X-1\atop \cos\pi y/L=2Y-1}}  
\label{N scaled density matrix}
\ee
and changing variables according to (\ref{change variables}) in
(\ref{ndm}) and (\ref{nC}) shows
\be
\rho_{N+1}^N(X,Y) = 
{\pi\rho\over\vert X -Y\vert^{1/2}}[X(1-X)]^{1/4}[Y(1-Y)]^{1/4}
{Z_{N,-1/2,-1/2}((X,1/2),(Y,1/2))\over
Z_{N+1,-1/2,-1/2}((\cdot,0),(\cdot,0))}.
\label{N partition ratio}
\ee
As already noticed in \cite{Forrester92,pra}, the log-gas
interpretation of (\ref{generalized partition}) allows us to predict
that for large $n$ it factorizes into a function of $X$ and the same
function of $Y$, which are themselves of the form (\ref{partition
definition}). Explicitly, we expect 
\be
{Z_{n,\lambda_1,\lambda_2}((X,q_1),(Y,q_2))\over Z_{n+q_1+q_2,\lambda_1,\lambda_2}((\cdot,0),(\cdot,0))}
\sim 
{Z_{n,\lambda_1,\lambda_2}((X,q_1))\over Z_{n+q_1,\lambda_1,\lambda_2}((\cdot,0))}
{Z_{n,\lambda_1,\lambda_2}((Y,q_2))\over Z_{n+q_2,\lambda_1,\lambda_2}((\cdot,0))}
\label{partition ratio asymptotics}
\ee
As in (\ref{partition definition}), the key to choosing the correct
normalizations is to balance the total charge in the log-gas
interpretation.
Setting  $q_1=q_2=q$ as required by (\ref{D partition ratio}) and
(\ref{N partition ratio}) it follows from (\ref{partition ratio
asymptotics}) that
\be
{Z_{n,\lambda_1,\lambda_2}((X,q),(Y,q))\over Z_{n+2q,\lambda_1,\lambda_2}((\cdot,0),(\cdot,0))}
\sim \left(
{1\over \pi^q}{G^2(q+1)\over G(2q+1)}(2n)^{-q+q^2}\right)^2
[X(1-X)]^{-q^2/2}[Y(1-Y)]^{-q^2/2}.
\ee 
Substituting this asymptotic form with $q=1/2$ in (\ref{D partition ratio}) and
(\ref{N partition ratio}) we obtain that for large $N$ and fixed
$X,Y\in(0,1)$,
\be
\rho_{N+1}^D(X,Y)\sim \rho_{N+1}^N(X,Y)\sim \rho {G^4(3/2)\over
\sqrt{2N}}
{[X(1-X)]^{1/8}[Y(1-Y)]^{1/8}\over \vert X-Y \vert^{1/2}}.
\label{density matrix asymptotics}
\ee

It is of interest to compare the asymptotic formula (\ref{density
matrix asymptotics}) against a numerical determination of say
$\rho_{N+1}^D(X,Y)$ or more conveniently $\rho_{N+1}^D(X,1-X)$. To
compute the latter we write it as a random matrix average. Thus it
follows from the various definitions that
\be
\rho_{N+1}^D(X,1-X)=
{8\rho\over N+1} X(1-X)\left<
\prod_{l=1}^N\left(\vert 4(1-X)-4X_l\vert\right)
\left(\vert 4X-4X_l\vert\right)
\right>_{{\textnormal J}{\textnormal U}{\textnormal E}_N}{\Bigg\vert}_{\lambda_1=\lambda_2=1/2}
\ee
For each $k=1,2,...M$ suppose we sample from
$\textnormal{J}\textnormal{U}\textnormal{E}_N\vert_{\lambda_1=\lambda_2=1/2}$ obtaining the $N$-tuple
$(X_1^{(k)},X_2^{(k)},...,X_N^{(k)})$. Then the method of Monte Carlo
integration tells us that
\ba
\rho_{N+1}^D(X,1-X)
&=&
{8\rho\over N+1}X(1-X){1\over M}\sum_{k=1}^M\prod_{l=1}^N\left(\vert 4(1-X)-4X_l^{(k)}\vert\right)
\left(\vert 4X-4X_l^{(k)}\vert\right)
\nonumber \\
&+&O\left(\frac{1}{\sqrt{M}}\right)
.
\label{sample}
\ea
Fortuitously, we have available a recently discovered
\cite{ForresterRains02b} random three term recurrence which generates
a polynomial, the zeros of which have the \pdf  $\textnormal{J}(2\lambda)\textnormal{E}_n$. In
the case of interest $(\lambda_1=\lambda_2=1/2,\lambda=1)$ the
recurrence states 
\ba
A_0(x)
&=&1
\nonumber \\
A_1(x)&=&x-\textnormal{B}\textnormal{D}[n+1/2,n+1/2]
\nonumber \\
A_j(x)&=&(w_2(x-1)+w_0x)A_{j-1}(x)+w_1x(x-1)A_{j-2}(x)\qquad (j=2,...,n)
\label{A recurrence}
\ea
where with 
\be
a\in \textnormal{G}\textnormal{D}[n+1-j+1/2,1], \quad b\in \textnormal{G}\textnormal{D}[j-1,1/2],\quad c\in
\textnormal{G}\textnormal{D}[n+1-j+1/2,1]
\ee
and $d:=a+b+c$ we have
\be 
w_0 ={a\over d}\qquad w_1={b\over d},\qquad w_2=1-w_0-w_1.
\ee
Here $\textnormal{B}\textnormal{D}[a,b]$ denotes the classical beta distribution, while
$\textnormal{G}\textnormal{D}[m,\sigma]$ denotes the classical  gamma distribution. The theory
of \cite{ForresterRains02b} tells us that $A_n(x)$ has its zeros
distributed according to
$\textnormal{J}\textnormal{U}\textnormal{E}_n\vert_{\lambda_1=\lambda_2=\lambda}$. Implementing (\ref{A
recurrence}) for fixed $n$ we thus computed the samples required in
(\ref{sample}) for the Monte Carlo evaluation of
$\rho_{N+1}^D(X,1-X)$. Forming the ratio then with the asymptotic form
(\ref{density matrix asymptotics}) gave the data in Table 
\ref{monte carlo}.
\begin{table*}
\begin{tabular}{|c|c|} \hline
$X$ &  $\rho_{N+1}^{D,MC}(X,1-X)/\rho_{N+1}^{D}(X,1-X)$\\
\hline\hline
$0.025$ &  $1.0958$                \\
\hline
$0.075$ &  $1.0039$               \\
\hline
$0.125$ &  $1.0363$               \\
\hline
$0.175$ &  $1.0098$           \\
\hline
$0.225$ &  $0.9439$                \\
\hline
$0.275$ &  $1.0080$               \\
\hline
$0.325$ &  $0.9692$               \\
\hline
$0.375$ &  $1.0338$           \\
\hline
$0.425$ &  $0.9706$                \\
\hline
$0.475$ &  $1.1309$               \\
\hline
\hline
\end{tabular}
\caption{The ratio $\rho_{N+1}^{D,MC}(X,1-X)/\rho_{N+1}^{D}(X,1-X)$
where $\rho_{N+1}^{D,MC}(X,1-X)$ refers to the Monte Carlo expression
(\ref{sample}), while $\rho_{N+1}^{D}(X,1-X)$ is the asymptotic form
(\ref{density matrix asymptotics}). We chose $N=14$, and evaluated (\ref{sample}) with $M=5000$.}
\label{monte carlo}
\end{table*}
\section{Physical and mathematical implications}
\label{implications}
\subsection{Ground state occupation of effective single particle
states}
The ground state density matrix is the theoretical quantity which
quantifies the condensation of a Bose system. Thus if we decompose the
density matrix
\be
\rho_N(x,y)=\sum_{j=0}^{\infty}\lambda_j\phi_j(x)\phi_j(y),
\label{density matrix expansion}
\ee
where the $\lambda_j,\phi_j$ are the eigenvalues and normalized
eigenfunctions in the eigenvalue problem
\be
\int\rho_N(x,y)\phi_j(y)\,dy=\lambda_j\phi_j(x),
\label{integral equation}
\ee
then by analogy with the free Fermi system in which (\ref{density
matrix expansion}) holds with {$\lambda_j=1$ ~$(j=0,...,N-1)$},
{$\lambda_j=0$ ~$(j\ge N)$}, we see that the $\lambda_j$ have the
physical interpretation as the occupation numbers of effective single
particle states $\phi_j(x)$. For Bose-Einstein condensation to occur
we must have $\lambda_0$ proportional to $N$.

To study (\ref{integral equation}) in the case of the impenetrable
Bose gas in Dirichlet or Neumann boundary conditions we restrict
ourselves to large $N$ where use can be made of the asymptotic form of
the density matrices (\ref{density matrix asymptotics}). Before making
the substitution, recalling the definitions (\ref{D scaled density
matrix}) and (\ref{N scaled density matrix}) we must first change
variables in (\ref{integral equation}) and redefine the eigenfunctions
so that 
\be
\cos\pi y/L= 2Y-1, \quad \phi_j(Y) = \phi_j(y)\vert_{\cos\pi y/L=
2Y-1}
, \quad \phi_j(X)=\phi_j(x)\vert_{\cos\pi x/L=2X-1}.
\label{scaled phi}
\ee
Doing this we obtain, for large $N$, the integral equation 
\be
\sqrt{N\over 2}{G^4(3/2)\over\pi}\int_0^1
{[X(1-X)]^{1/8}[Y(1-Y)]^{1/8}\over \vert X-Y\vert^{1/2}}
\phi_j(Y)
{dY\over \sqrt{Y(1-Y)}}
=\lambda_j\phi_j(X).
\label{scaled integral equation}
\ee
It follows immediately that 
\be
\lambda_j\propto \sqrt{N}.
\label{root N}
\ee
As noted in \cite{pra} this conclusion requires that $j$ be fixed \---
for $j\gg N$ we expect $\lambda_j\propto (N/j)^4$ in keeping with the
corresponding result in periodic boundary conditions, since in this
regime the boundary conditions are not expected to play a role.

Setting 
\be
\lambda_j={G^4(3/2)\over \sqrt{2}\pi}\sqrt{N}{\bar \lambda}_j
\label{scaled eigenvalue definition}
\ee
and rearranging, (\ref{scaled integral equation}) reads
\be
\int_0^1{\phi_j(Y)\over\vert X-Y\vert^{1/2}}{dY\over [Y(1-Y)]^{3/8}}
={\bar \lambda}_j {\phi_j(X)\over [X(1-X)]^{1/8}}.
\label{norm form integral equation}
\ee
Remarkably the effective single particle ground state $\phi_0(X)$, and
the corresponding scaled occupation number ${\bar \lambda}_0$ can be
computed exactly from (\ref{norm form integral equation}). To see this
requires knowledge of a piece of integral equation theory presented in
Porter and Stirling \cite{PorterStirling}. The relevant theory tells
us that the solution of the integral equation
\be
\int_0^1{\phi(t)\over\vert x-t\vert^{\nu}}dt=1, \quad\qquad \nu<1
\label{PorterStirling equation}
\ee
is 
\be 
\phi(x)={1\over\pi}\left(\cos{\pi\nu\over 2}\right)[x(1-x)]^{(\nu-1)/2}.
\label{PorterStirling solution}
\ee
Setting $\nu=1/2$, it follows immediately that 
\be
\phi_0(X)={1\over\sqrt{A}}[X(1-X)]^{1/8},\qquad
{\bar\lambda}_0=\pi\sqrt{2}
\label{ground state}
\ee
satisfies (\ref{PorterStirling equation}), where the normalization $A$
is determined by the requirement that
\be
{L\over\pi}
\int_0^1(\phi_0(X))^2{dX\over\sqrt{X(1-X)}}=1,
\label{ground state normalization}
\ee
and so 
\be
A={L\over\pi}B(3/4,3/4)
\ee
where $B(a,b)$ denotes the beta function. Substituting the exact
evaluation of ${\bar\lambda}_0$ in (\ref{scaled eigenvalue definition}) shows
that in the large $N$ limit
\be
\lambda_0=G^4(3/2)\sqrt{N}=1.3069\sqrt{N}.
\ee

To compute the higher order single particle states and their
corresponding occupations we make the ansatz
\be
\phi_j(X)\propto\phi_0(X)p_j(X)
\label{product ansatz}
\ee  
where $p_j(X)$ is a polynomial of degree $j$. Now $\{\phi_j(X)\}$ can
always be chosen to be orthogonal (note that the measure is
$dX/\sqrt{X(1-X)}$ on $[0,1]$) so recalling (\ref{ground state})
 we require
\be
\int_0^1\frac{p_j(X)p_k(X)}{(X(1-X))^{1/4}}dX
=0
\qquad j\ne k .
\ee
Up to normalization, the unique polynomials with this property are the
particular Gegenbauer polynomials
\be
p_j(X)=C_j^{1/4}(2X-1),
\label{gegenbauer}
\ee
which we note are proportional to the particular Jacobi polynomials $P^{-1/4,-1/4}_j(2X-1)$.
Normalizing (\ref{product ansatz}) with the substitution (\ref{gegenbauer})
as in (\ref{ground state normalization}) shows
\be
\phi_j(X)=\sqrt{\frac{1}{L}}\sqrt{\frac{j!(j+1/4)\Gamma^2(1/4)}{\Gamma(j+1/2)}}
(X(1-X))^{1/8}C_j^{1/4}(2X-1).
\label{excited states}
\ee
Now substituting (\ref{excited states}) in (\ref{norm form integral
  equation}) and setting $X=1$ we obtain a definite integral for
${\bar\lambda}_j$ which can be found in  \cite{Gradstein}, giving us the
evaluation
\be
{\bar\lambda}_j=\sqrt{2\pi}\frac{\Gamma(j+1/2)}{j!}
\label{scaled eigenvalue}
\ee
and hence 
\be
\lambda_j=G^4(3/2)\frac{\Gamma(j+1/2)}{\sqrt{\pi}j!}\sqrt{N}.
\ee

To arrive at (\ref{excited states}) we have made the ansatz
(\ref{product ansatz}). In fact a different approach can be taken to
the problem, in which it is shown that an integral operator following
from (\ref{norm form integral equation}) commutes with the
differential operator determining the polynomials
$\{C_j^{1/4}(2X-1)\}_{j=0,1,2,...}$.This is done in Appendix \ref{proof}.

Finally, we note that substituting (\ref{density matrix asymptotics}), (\ref{excited states}) 
and (\ref{scaled eigenvalue}) in (\ref{density matrix expansion})
gives the following interesting identity
\be
\frac{1}{\vert X-Y\vert^{1/2}}
=
\sqrt{\frac{2}{\pi}}\Gamma^2(1/4)
\sum_{j=0}^{\infty}(j+1/4)C_j^{1/4}(2X-1)C_j^{1/4}(2Y-1).
\ee  

\subsection{Generalized Fisher-Hartwig type asymptotics}
One viewpoint of our asymptotic analysis of multiple integrals of the
form (\ref{generalized partition}) is that we are studying asymptotic
problems of the Fisher-Hartwig class. Let us recall that the latter
refers literally to Toeplitz determinants with both zeros and jump
discontinuities is its generating function,
\be
D_N[e^{a(\theta)}]:=\det[a_{i-j}]_{i,j=1,...,N},
\qquad e^{a(\theta)}=\sum_{p=-\infty}^{\infty}a_pe^{ip\theta}
\ee
where
\be
a(\theta)=g(\theta)-i\sum_{r=1}^Rb_r[\pi-(\theta-\phi_r)]{\text{\text{\text{mod}}}} 2\pi 
+\sum_{r=1}^Ra_r\log\vert 2-2\cos(\theta -\phi_r)\vert
\ee
with 
\be
g(\theta)=\sum_{p=-\infty}^{\infty}g_pe^{ip\theta},\qquad\quad
\sum_{p=-\infty}^{\infty}\vert p\vert \vert g_p\vert^2<\infty.
\ee
Thus $g(\theta)$ is a regular term, while at $\phi_r$ $(r=1,...,R)$
there is a jump discontinuity of strength $b_r$ and a zero of order
$a_r$. To see the relationship with (\ref{generalized partition}) we
set $b_r=0$ $(r=1,...,R)$ thus eliminating the jump discontinuities,
and recall the general formula relating a Toeplitz determinant to a
multiple integral,
\ba
D_N[e^{a(\theta)}]&=&
{1\over N!}\int_0^{2\pi}d\theta_1...\int_0^{2\pi}d\theta_N
\prod_{l=1}^N e^{a(\theta_l)}\prod_{1\le j<k\le N}\vert e^{i\theta_k}-e^{i\theta_j}\vert^2 
\nonumber \\
&=&
{1\over N!}\int_0^{2\pi}d\theta_1...\int_0^{2\pi}d\theta_N\prod_{l=1}^N e^{g(\theta_l)}
\left(\prod_{r=1}^R\vert
e^{i\theta_l}-e^{i\phi_r}\vert^{2a_r}\right)
\prod_{1\le j<k\le N}\vert e^{i\theta_k}-e^{i\theta_j}\vert^2.
\nonumber \\
&&
\label{FisherHartwig}
\ea

Fisher and Hartwig \cite{FisherHartwig68} conjectured that in the case
(\ref{FisherHartwig}),
\be
D_N[e^{a(\theta)}]\sim 
e^{g_0N}e^{\sum_{r=1}^Ra_r^2\log N}E
\label{FisherHartwig conjecture}
\ee
where $E$ is independent of $N$. This was subsequently proved, and it
was furthermore shown 
\be
E
=
e^{\sum_{k=1}^{\infty}kg_kg_{-k}}\prod_{r=1}^Re^{-a_r(g(\phi_r)-g_0)}
\prod_{1\le j<k\le R}\vert e^{i\phi_k}-e^{i\phi_j}\vert^{-2a_ka_j}
\prod_{r=1}^R{G^2(1+a_r)\over G(1+2a_r)}
\label{E}
\ee
(see for example the monograph \cite{BottcherSilbermann89} and references
therein).

As noted by one of the present authors some years ago
\cite{ForresterPisani}, it is straight forward to reproduce the
structure of (\ref{FisherHartwig conjecture}) using the analogous
log-gas argument to that used here in the analysis of
(\ref{generalized partition}). Now the (normalized) multiple integral
corresponding to (\ref{FisherHartwig}) which relates to
(\ref{generalized partition}) is 
\be
{H_{n,\lambda_1,\lambda_2}[e^{h(x)}\prod_{r=1}^R\vert y_r
-x\vert^{2q_r}]
\over
H_{n,\lambda_1,\lambda_2}[1]},
\label{H ratio}
\ee
\be
H_{n,\lambda_1,\lambda_2}[f(x)]:=\int_0^1dx_1...\int_0^1dx_n
\prod_{l=1}^nf(x_l) x_l^{\lambda_1}(1-x_l)^{\lambda_2}
\prod_{1\le j<k\le n}\vert x_k -x_j\vert^2
\label{H definition}
\ee
where $h(x)$ is analytic on $(0,1)$. As our final issue, we would like to extend the
log-gas argument used in the analysis of (\ref{generalized partition})
to predict the large $n$ asymptotic form of (\ref{H ratio}). 

From the log-gas perspective, the natural quantity to analyze is 
\be
\prod_{1\le j<k\le r}\vert y_k - y_j\vert^{2q_jq_k}
{H_{n,\lambda_1,\lambda_2}[e^{h(x)}\prod_{r=1}^R\vert y_r
-x\vert^{2q_r}]
\over
H_{n+\sum_{r=1}^Rq_r,\lambda_1,\lambda_2}[e^{h(x)}]}
\label{log-gas natural quantity}
\ee
where for $m$ a non-negative integer
\be
H_m[h]:=\int_0^1dx_1...\int_0^1dx_m\prod_{l=1}^m
e^{h(x_l)}x_l^{\lambda_1}(1-x_l)^{\lambda_2}
\prod_{1\le j<k \le m}
\vert x_k -x_j \vert^2.
\ee
Analogous to (\ref{partition ratio asymptotics}) we expect for large
$n$ (\ref{log-gas natural quantity}) to factorize as 
\be
{H_{n,\lambda_1,\lambda_2}[e^{h(x)}\prod_{r=1}^R\vert y_r-x\vert^{2q_r}]\over
H_{n+\sum_{r=1}^Rq_r,\lambda_1,\lambda_2}[e^{h(x)}]}\sim
\prod_{r=1}^Re^{-q_rh(y_r)}
{H_{n,\lambda_1,\lambda_2}[\vert y_r-x\vert^{2q_r}]\over
H_{n+q_r,\lambda_1,\lambda_2}[1]}
\label{H factorization}
\ee
where the second expression is motivated by inspection of the known
results (\ref{FisherHartwig conjecture}) and (\ref{E}) for
(\ref{FisherHartwig}). 

Thus we expect 
\ba
{H_{n,\lambda_1,\lambda_2}[e^{h(x)}\prod_{r=1}^R\vert y_r-x\vert^{2q_r}]
\over H_{n,\lambda_1,\lambda_2}[1]}
&\sim&
\prod_{1\le j<k\le r}\vert y_k -y_j\vert^{-2q_jq_k}
{H_{n+\sum_{r=1}^Rq_r,\lambda_1,\lambda_2}[e^{h(x)}]\over
H_{n,\lambda_1,\lambda_2}[1]}
\nonumber \\
&\times&
\prod_{r=1}^R e^{-q_rh(y_r)}{H_{n,\lambda_1,\lambda_2}[\vert
y_r-x\vert^{2q_r}]
\over H_{n+q_r,\lambda_1,\lambda_2}[1]}.
\ea
But according to (\ref{asymptotic partition ratio})
\be
{H_{n,\lambda_1,\lambda_2}[\vert y_r-x\vert^{2q_r}]\over
H_{n+q_r,\lambda_1,\lambda_2}[1]}
\sim
{1\over \pi^{q_r}}
{G^2(q_r+1)\over G(2q_r +1)}
(2n)^{-q_r+q_r^2}
(y_r(1-y_r))^{-q_r^2/2}.
\label{asymptotic H ratio 1}
\ee
Also, for the first ratio in (\ref{H factorization}) we have available
both rigorous results \cite{Hirschmann65,Johansson98} as well as
log-gas type arguments \cite{Beenakker93} which together tell us that
\ba
{H_{n+Q,\lambda_1,\lambda_2}[e^{h(x)}]\over
H_{n,\lambda_1,\lambda_2}[1]}
&\sim&
\exp\left[{n+Q+(\lambda_1+\lambda_2)/2\over \pi}\int_0^1{h(x)\over
[x(1-x)]^{1/2}}dx\right]
\nonumber \\
&\times &
\exp\left[-{\lambda_1+\lambda_2\over 4}(h(0)+h(1))
\right]
\nonumber \\
&\times&
\exp\left[
{1\over 4\pi^2}
\int_0^1dx{h(x)\over [x(1-x)]^{1/2}}\int_0^1dy
{h'(y)[y(1-y)]^{1/2}\over x-y}\right].
\label{asymptotic H ratio 2}
\ea

Substituting (\ref{asymptotic H ratio 1}) and (\ref{asymptotic H ratio
2}) in (\ref{H factorization}) gives the analogue of
(\ref{FisherHartwig conjecture}),
\ba
{H_{n,\lambda_1,\lambda_2}[e^{h(x)}\prod_{r=1}^R\vert y_r -x\vert^{2q_r}]
\over H_{n,\lambda_1,\lambda_2}[1]}
&\sim&
\exp\left[{n+\sum_{r=1}^Rq_r+(\lambda_1+\lambda_2)/2\over
\pi}\int_0^1{h(x)\over[x(1-x)]^{1/2}}dx
\right]
\nonumber \\
&\times&
\exp\left[\sum_{r=1}^R(-q_r+q_r^2)\log 2n
\right] K
\label{Jacobi Fisher Hartwig}
\ea
where
\ba
K
&=&
\prod_{1\le j<k \le R}\vert y_k-y_j\vert^{-2q_jq_k}e^{-(\lambda_1+\lambda_2)[h(0)+h(1)]/4}
e^{-\sum_{r=1}^R q_rh(y_r)}
\nonumber \\
&\times& 
\exp\left[{1\over 4\pi^2}\int_0^1dx {h(x)\over [x(1-x)]^{1/2}}
\int_0^1dy{h'(y)[y(1-y)]^{1/2}\over x-y}\right]\prod_{r=1}^R[y_r(1-y_r)]^{-q_r^2/2}
\nonumber \\
&\times&
\prod_{r=1}^R{1\over\pi^{q_r}}
{G^2(q_r+1)\over G(2q_r+1)}.
\label{K}
\ea
\subsection{Concluding remarks}
In our first paper on the impenetrable Bose gas \cite{cmp} we set
ourselves the goal of providing the leading asymptotic form of the
density matrix for the impenetrable Bose gas in a harmonic trap and in
Dirichlet and Neumann boundary conditions. It was noted in \cite{cmp}
that for the impenetrable Bose gas in periodic boundary conditions,
the Fisher-Hartwig formula gave the asymptotic form
\be
\rho_{N+1}^C(x;0)\sim \rho \frac{G^4(3/2)}{\sqrt{2\pi}}
\left(\frac{\pi}{N\sin(\pi\rho x/N)}\right)^{1/2}.
\ee
In \cite{pra}, it was shown that for the harmonic well 
\be
(2N)^{1/2}\rho^H_{N+1}(\sqrt{2N}X,\sqrt{2N}Y)\sim
N^{1/2}\frac{G^4(3/2)}{\pi}
\frac{(1-X^2)^{1/8}(1-Y^2)^{1/8}}{\vert X-Y\vert^{1/2}},
\ee
while in the present paper, after changing variables according to
(\ref{D scaled density matrix}), the asymptotic form of the density matrix is shown to
have the leading asymptotic form (\ref{density matrix asymptotics}).
As a consequence of the scaling properties of these asymptotic forms,
the occupation number of the low lying effective single particle
states are all proportional to $\sqrt{N}$, but with a proportionality
constant dependent on the particular system.

To obtain the asymptotic forms we have used a combination of exact
analysis, made possible by the theory of Selberg correlation
integrals, and physical reasoning based on log-gas analogies. Taking
this argument to its logical conclusion leads to a conjectured exact
asymptotic formula, given by (\ref{Jacobi Fisher Hartwig}) and (\ref{K}) for a Jacobi weight
analogue of the Fisher-Hartwig formula. 
\appendix
\section{Proof that $\{\phi_0(X) C_j^{1/4}(2X-1)\}_{j=0,1,2,...}$ are
    solutions of the integral equation (\ref{norm form integral equation})}
\label{proof}
The assertion that the $\phi_j(X)$ given by (\ref{product ansatz}) are
solutions of (\ref{norm form integral equation}) is equivalent to
stating that the Gegenbauer polynomials are  eigenfunctions of the integral operator 
\be
K[f(\xi)]:=\int_{-1}^1\frac{d\psi}{\vert \xi-\psi\vert^{1/2}(1-\psi^2)^{1/4}}f(\psi),
\ee
where for convenience we are working on the interval $-1\le\xi\le 1$.
In this Appendix we prove that $K$ commutes with the
differential operator, $L$, which determines the Gegenbauer polynomials
\be
L:= (1-\xi^2)\frac{d^2}{d\xi^2}-\frac{3}{2}\xi\frac{d}{d\xi},
\ee
with
\be
LC_j^{1/4}(\xi)=-j(j+1/2)C_j^{1/4}(\xi).
\ee

We begin by identifying $K[\C(\xi)]$ with the following finite sum of hypergeometric
functions
\ba
K[\C(\xi)]
=
&\Omega_j& \sum_{k=0}^j
\frac{(-j)_k(j+1/2)_k}{k! (3/4)_k}
\left(\frac{1+\xi}{2}\right)^{1/4}
{_2F_1}\left(\frac{1}{4}-k,\frac{3}{4};\frac{5}{4};\frac{1+\xi}{2}\right)
\nonumber \\
+(-1)^j
&\Omega_j& \sum_{k=0}^j
\frac{(-j)_k(j+1/2)_k}{k! (3/4)_k}
\left(\frac{1-\xi}{2}\right)^{1/4}
{_2F_1}\left(\frac{1}{4}-k,\frac{3}{4};\frac{5}{4};\frac{1-\xi}{2}\right),
\label{hypergeometric representation}
\ea
where 
\be
\Omega_j:= \frac{\Gamma(3/4)}{\Gamma(5/4)}\frac{\Gamma(j+1/2)}{j!}.
\ee
To derive (\ref{hypergeometric representation}) we break up the  interval of integration in  $K[\C(\xi)]$
into two regions, thus removing the need for the modulus, which gives
\ba
K[\C(\xi)]
&=&
\int_{\xi}^1(1-\psi^2)^{-1/4}\C(\psi)\frac{d\psi}{\sqrt{\psi-\xi}}
+(-1)^j\int_{-\xi}^1(1-\psi^2)^{-1/4}\C(\psi)\frac{d\psi}{\sqrt{\psi+\xi}}
\nonumber \\
&=&
z_-^{1/4}\int_0^1d\psi \,\psi^{-1/4}(1-\psi)^{-1/2}
\left(1-z_-\psi\right)^{-1/4}\C(1-2z_-\psi)
\nonumber \\
+&(-1)^j&
z_+^{1/4}\int_0^1d\psi \,\psi^{-1/4}(1-\psi)^{-1/2}
\left(1-z_+\psi\right)^{-1/4}\C(1-2z_+\psi)
\label{break up}
\ea
where we have defined 
\be
z_{\pm}:=\frac{1\pm \xi}{2}
\ee
and in obtaining the second equality we have changed 
integration variables from $\psi$ to \newline
{$(1-\psi)/(1-\xi)$} and
{$(1-\psi)/(1+\xi)$} in the first and second integrals respectively. 
Substituting the following known \cite{Gradstein} expansion for
the Gegenbauer polynomials 
\be
\C(\psi)=\frac{(-1)^j\Gamma(j+1/2)}{j!\sqrt{\pi}}
\sum_{k=0}^j
\frac{(-j)_k(j+1/2)_k}{k!(3/4)_k}\left(\frac{1+\psi}{2}\right)^k
\ee
into (\ref{break up}), and using the standard integral representation of the $_2F_1$ function
\be
\int_0^1
d\psi
\frac{\psi^{-1/4}(1-\psi)^{-1/2}}{(1-z\psi)^{-k+1/4}}=
B(3/4,1/2)
{_2F_1\left(\frac{1}{4}-k,\frac{3}{4};\frac{5}{4};z\right)},
\ee
then results in (\ref{hypergeometric representation}). 

We can now utilise known hypergeometric identities to facilitate the
operation of the differential operator $L$ on  the expression for
$K[\C(\xi)]$ given by
(\ref{hypergeometric representation}).
We note that the structure of
(\ref{hypergeometric representation}) is of the form
\be
K[\C(\xi)]=\Omega_jS_j(z_+)+(-1)^j\Omega_jS_j(z_-),
\ee
where 
\be
S_j(z):= \sum_{k=0}^j \frac{(-j)_k(j+1/2)_k}{k!(3/4)_k}
z^{1/4}{_2F_1}\left(\frac{1}{4}-k,\frac{3}{4};\frac{5}{4};z\right).
\label{S definition}
\ee
In terms of both the variables $z=z_+$ and $z=z_-$ the differential operator $L$ has
the form 
\be
L=z(1-z)\frac{d^2}{dz^2}
-\frac{3}{4}(2z-1)\frac{d}{dz}.
\ee
Utilising the identity \cite{Luke}
\be
\frac{d^n}{dz^n}\left[z^{\delta}{_pF_q}\left(
{\alpha_1,...,\alpha_p \atop\rho_1,...,\rho_q}{\Big\vert}z\right)\right]
=(\delta-n+1)_n z^{\delta-n}{_{p+1}F_{q+1}}
\left({\delta+1,\alpha_1,...,\alpha_p \atop \delta+1-n,\rho_1,...,\rho_q}{\Big\vert}z\right)
\label{luke differentiation}
\ee
we find that
\ba
Lz^{1/4}{_2F_1}\left(\frac{1}{4}-k,\frac{3}{4};\frac{5}{4};z\right)
&=&
\frac{z^{1/4}}{z}\left(-\frac{3}{16}\right)(1-z)\,{_2F_1}\left(\frac{1}{4}-k,\frac{3}{4};-\frac{3}{4};z\right)
\nonumber \\
&+&
\frac{z^{1/4}}{z}\left(-\frac{3}{16}\right)(2z-1)\,{_2F_1}\left(\frac{1}{4}-k,\frac{3}{4};\frac{1}{4};z\right)
\label{differentiation}
\ea
\ba
Lz^{1/4}{_2F_1}\left(\frac{1}{4}-k,\frac{3}{4};\frac{5}{4};z\right)
&=&
-\frac{1}{2}z^{1/4}k\;{_2F_1}\left(\frac{1}{4}-k,\frac{3}{4};\frac{5}{4};z\right)
\nonumber \\
&&-\frac{1}{4}
z^{1/4}k\;{_2F_1}\left(\frac{1}{4}-k,\frac{3}{4};\frac{1}{4};z\right)
\label{contiguity 1}
\\Lz^{1/4}{_2F_1}\left(\frac{1}{4}-k,\frac{3}{4};\frac{5}{4};z\right)
&=&
-z^{1/4}k\left(k+\frac{1}{2}\right){_2F_1}\left(\frac{1}{4}-k,\frac{3}{4};\frac{5}{4};z\right)
\nonumber \\
&&-
z^{1/4}k\left(\frac{1}{4}-k\right){_2F_1}\left(\frac{1}{4}-(k-1),\frac{3}{4};\frac{5}{4};z\right)
\label{contiguity 2}
\ea
where the equalities in (\ref{contiguity 1}) and (\ref{contiguity 2}) respectively
follow from the two particular contiguity relations 
\cite{Luke}
\ba
\left(-\frac{3}{16}\right)
 (1-z){_2F_1}\left(\frac{1}{4}-k,\frac{3}{4};-\frac{3}{4};z\right)
&=&
\frac{1}{16}\left[(6-4k)z-3\right] {_2F_1}\left(\frac{1}{4}-k,\frac{3}{4};\frac{1}{4};z\right)
\nonumber \\ 
&-&
\frac{k z}{2}{_2F_1}\left(\frac{1}{4}-k,\frac{3}{4};\frac{5}{4};z\right),
\label{contiguity 1 lemma}
\ea
and
\be
-\frac{1}{4}
 {_2F_1}\left(\frac{1}{4}-k,\frac{3}{4};\frac{1}{4};z\right)
 =-k\, {_2F_1}\left(\frac{1}{4}-k,\frac{3}{4};\frac{5}{4};z\right)
 -\left(\frac{1}{4}-k\right){_2F_1}\left(\frac{1}{4}-(k-1),\frac{3}{4};\frac{5}{4};z\right).
\label{contiguity 2 lemma}
\ee
Hence from (\ref{contiguity 2})  we obtain
\ba
LS_j(z)
&=&
-z^{1/4}\sum_{k=0}^j
\frac{(-j)_k(j+1/2)_k}{k!(3/4)_k}
k\left(\frac{1}{2}+k\right){_2F_1}\left(\frac{1}{4}-k,\frac{3}{4};\frac{5}{4};z\right)
\nonumber \\
&&-
z^{1/4}\sum_{k=0}^j
\frac{(-j)_k(j+1/2)_k}{k!(3/4)_k}
k\left(\frac{1}{4}-k\right){_2F_1}\left(\frac{1}{4}-(k-1),\frac{3}{4};\frac{5}{4};z\right).
\label{S before change of index}
\ea
Changing summation index in the second sum in (\ref{S before change of
  index}) from $k$ to $k-1$ and simplifying we deduce finally that 
\ba
LS_j(z)&=&-j(j+1/2)\sum_{k=0}^j
\frac{(-j)_k(j+1/2)_k}{k!(3/4)_k}
z^{1/4}{_2F_1}\left(\frac{1}{4}-k,\frac{3}{4};\frac{5}{4};z\right)
\nonumber \\
&=&-j(j+1/2)S_j(z),
\ea
which then implies that 
\ba
L K C^{1/4}_j(\xi)
&=&
-j(j+1/2)KC^{1/4}_j(\xi)
\\
&=&
K L C^{1/4}_j(\xi) \qquad\qquad\qquad (j=0,1,2,...)
\ea
and so
\be
[K,L]=0.
\ee

\begin{acknowledgments}
The authors would like to thank the Australian Research Council for
funding this work. 
\end{acknowledgments}


\begin{thebibliography}{32}
\expandafter\ifx\csname natexlab\endcsname\relax\def\natexlab#1{#1}\fi
\expandafter\ifx\csname bibnamefont\endcsname\relax
  \def\bibnamefont#1{#1}\fi
\expandafter\ifx\csname bibfnamefont\endcsname\relax
  \def\bibfnamefont#1{#1}\fi
\expandafter\ifx\csname citenamefont\endcsname\relax
  \def\citenamefont#1{#1}\fi
\expandafter\ifx\csname url\endcsname\relax
  \def\url#1{\texttt{#1}}\fi
\expandafter\ifx\csname urlprefix\endcsname\relax\def\urlprefix{URL }\fi
\providecommand{\bibinfo}[2]{#2}
\providecommand{\eprint}[2][]{\url{#2}}

\bibitem[{\citenamefont{Forrester}()}]{ForresterBook}
\bibinfo{author}{\bibfnamefont{P.~J.} \bibnamefont{Forrester}},
  \emph{\bibinfo{title}{Log gases and random matrices}},
  \urlprefix\url{http://www.ms.unimelb.edu.au/~matpjf/matpjf.html}.

\bibitem[{\citenamefont{Forrester et~al.}()\citenamefont{Forrester, Frankel,
  Garoni, and Witte}}]{cmp}
\bibinfo{author}{\bibfnamefont{P.~J.} \bibnamefont{Forrester}},
  \bibinfo{author}{\bibfnamefont{N.~E.} \bibnamefont{Frankel}},
  \bibinfo{author}{\bibfnamefont{T.~M.} \bibnamefont{Garoni}},
  \bibnamefont{and} \bibinfo{author}{\bibfnamefont{N.~S.} \bibnamefont{Witte}},
  \bibinfo{note}{in press Commun. Math. Phys.}, \eprint{math-ph/0207005}.

\bibitem[{\citenamefont{Katz and Sarnak}(1999)}]{KatzSarnakBook}
\bibinfo{author}{\bibfnamefont{N.~M.} \bibnamefont{Katz}} \bibnamefont{and}
  \bibinfo{author}{\bibfnamefont{P.}~\bibnamefont{Sarnak}},
  \emph{\bibinfo{title}{Random Matrices, Frobenius Eigenvalues and Monodromy}}
  (\bibinfo{publisher}{AMS}, \bibinfo{address}{Providence, Rhode Island},
  \bibinfo{year}{1999}).

\bibitem[{\citenamefont{Keating and Snaith}(2000)}]{KeatingSnaithPaper}
\bibinfo{author}{\bibfnamefont{J.~P.} \bibnamefont{Keating}} \bibnamefont{and}
  \bibinfo{author}{\bibfnamefont{N.~C.} \bibnamefont{Snaith}},
  \bibinfo{journal}{Commum. Math. Phys.} \textbf{\bibinfo{volume}{214}},
  \bibinfo{pages}{91} (\bibinfo{year}{2000}).

\bibitem[{\citenamefont{Keating and Snaith}()}]{KeatingSnaithReview03}
\bibinfo{author}{\bibfnamefont{J.~P.} \bibnamefont{Keating}} \bibnamefont{and}
  \bibinfo{author}{\bibfnamefont{N.~C.} \bibnamefont{Snaith}},
  \bibinfo{note}{to appear J. Phys. A.}

\bibitem[{\citenamefont{Forrester}(1992)}]{Forrester92}
\bibinfo{author}{\bibfnamefont{P.~J.} \bibnamefont{Forrester}},
  \bibinfo{journal}{Phys. Lett. A} \textbf{\bibinfo{volume}{163}},
  \bibinfo{pages}{121} (\bibinfo{year}{1992}).

\bibitem[{\citenamefont{Forrester et~al.}(2003)\citenamefont{Forrester,
  Frankel, Garoni, and Witte}}]{pra}
\bibinfo{author}{\bibfnamefont{P.~J.} \bibnamefont{Forrester}},
  \bibinfo{author}{\bibfnamefont{N.~E.} \bibnamefont{Frankel}},
  \bibinfo{author}{\bibfnamefont{T.~M.} \bibnamefont{Garoni}},
  \bibnamefont{and} \bibinfo{author}{\bibfnamefont{N.~S.} \bibnamefont{Witte}},
  \bibinfo{journal}{Phys. Rev. A.} \textbf{\bibinfo{volume}{67}},
  \bibinfo{pages}{043607} (\bibinfo{year}{2003}).

\bibitem[{\citenamefont{Kaneko}(1993)}]{Kaneko93}
\bibinfo{author}{\bibfnamefont{J.}~\bibnamefont{Kaneko}},
  \bibinfo{journal}{SIAM J. Math. Anal.} \textbf{\bibinfo{volume}{24}},
  \bibinfo{pages}{1086} (\bibinfo{year}{1993}).

\bibitem[{\citenamefont{Macdonald}(1995)}]{Macdonald95}
\bibinfo{author}{\bibfnamefont{I.~G.} \bibnamefont{Macdonald}},
  \emph{\bibinfo{title}{Hall polynomials and symmetric functions}}
  (\bibinfo{publisher}{Oxford University Press}, \bibinfo{address}{Oxford},
  \bibinfo{year}{1995}), \bibinfo{edition}{2nd} ed.

\bibitem[{\citenamefont{Forrester}(1993{\natexlab{a}})}]{Forrester93a}
\bibinfo{author}{\bibfnamefont{P.~J.} \bibnamefont{Forrester}},
  \bibinfo{journal}{Phys. Lett. A.} \textbf{\bibinfo{volume}{179}},
  \bibinfo{pages}{127} (\bibinfo{year}{1993}{\natexlab{a}}).

\bibitem[{\citenamefont{Forrester}(1993{\natexlab{b}})}]{Forrester93b}
\bibinfo{author}{\bibfnamefont{P.~J.} \bibnamefont{Forrester}},
  \bibinfo{journal}{J. Math. Phys.} \textbf{\bibinfo{volume}{35}},
  \bibinfo{pages}{2539} (\bibinfo{year}{1993}{\natexlab{b}}).

\bibitem[{\citenamefont{Forrester}(1994)}]{Forrester94}
\bibinfo{author}{\bibfnamefont{P.~J.} \bibnamefont{Forrester}},
  \bibinfo{journal}{Nucl. Phys. B.} \textbf{\bibinfo{volume}{416}},
  \bibinfo{pages}{377} (\bibinfo{year}{1994}).

\bibitem[{\citenamefont{Baker and Forrester}(1997)}]{BakerForrester97}
\bibinfo{author}{\bibfnamefont{T.~H.} \bibnamefont{Baker}} \bibnamefont{and}
  \bibinfo{author}{\bibfnamefont{P.~J.} \bibnamefont{Forrester}},
  \bibinfo{journal}{Commun. Math. Phys.} \textbf{\bibinfo{volume}{188}},
  \bibinfo{pages}{1371} (\bibinfo{year}{1997}).

\bibitem[{\citenamefont{Forrester}(1998)}]{Forrester98}
\bibinfo{author}{\bibfnamefont{P.~J.} \bibnamefont{Forrester}}, in
  \emph{\bibinfo{booktitle}{Random Matrices, Log-gases and the
  Calogero-Sutherland Model, {\textnormal i{\textnormal n}} Quantum Many-Body
  Problems and Representation Theory}} (\bibinfo{publisher}{Mathematical
  Society of Japan}, \bibinfo{address}{Tokyo}, \bibinfo{year}{1998}),
  vol.~\bibinfo{volume}{1} of \emph{\bibinfo{series}{MSJ Memoirs}}, pp.
  \bibinfo{pages}{97--181}.

\bibitem[{\citenamefont{Forrester and Witte}()}]{ForresterWitte02}
\bibinfo{author}{\bibfnamefont{P.~J.} \bibnamefont{Forrester}}
  \bibnamefont{and} \bibinfo{author}{\bibfnamefont{N.~S.} \bibnamefont{Witte}},
  \eprint{math-ph/0204008}.

\bibitem[{\citenamefont{Forrester and Witte}(2001)}]{ForresterWitte01}
\bibinfo{author}{\bibfnamefont{P.~J.} \bibnamefont{Forrester}}
  \bibnamefont{and} \bibinfo{author}{\bibfnamefont{N.~S.} \bibnamefont{Witte}},
  \bibinfo{journal}{Commun. Math. Phys.} \textbf{\bibinfo{volume}{219}},
  \bibinfo{pages}{357} (\bibinfo{year}{2001}).

\bibitem[{\citenamefont{Brezin and Hikami}(2001)}]{BrezinHikami01}
\bibinfo{author}{\bibfnamefont{E.}~\bibnamefont{Brezin}} \bibnamefont{and}
  \bibinfo{author}{\bibfnamefont{S.}~\bibnamefont{Hikami}},
  \bibinfo{journal}{Commun. Math. Phys.} \textbf{\bibinfo{volume}{223}},
  \bibinfo{pages}{363} (\bibinfo{year}{2001}).

\bibitem[{\citenamefont{Mulase and Waldron}()}]{MulaseWaldron02}
\bibinfo{author}{\bibfnamefont{M.}~\bibnamefont{Mulase}} \bibnamefont{and}
  \bibinfo{author}{\bibfnamefont{A.}~\bibnamefont{Waldron}},
  \eprint{math-ph/0206011}.

\bibitem[{\citenamefont{Br\'ezin and Hikami}(2000)}]{BrezinHikami00}
\bibinfo{author}{\bibfnamefont{E.}~\bibnamefont{Br\'ezin}} \bibnamefont{and}
  \bibinfo{author}{\bibfnamefont{S.}~\bibnamefont{Hikami}},
  \bibinfo{journal}{Commun. Math. Phys.} \textbf{\bibinfo{volume}{214}},
  \bibinfo{pages}{111} (\bibinfo{year}{2000}).

\bibitem[{\citenamefont{Strahov and Fyodorov}()}]{StrahovFyodorov02}
\bibinfo{author}{\bibfnamefont{E.}~\bibnamefont{Strahov}} \bibnamefont{and}
  \bibinfo{author}{\bibfnamefont{Y.~V.} \bibnamefont{Fyodorov}},
  \eprint{math-ph/0210010}.

\bibitem[{\citenamefont{Mehta}(1991)}]{Mehta91}
\bibinfo{author}{\bibfnamefont{M.~L.} \bibnamefont{Mehta}},
  \emph{\bibinfo{title}{Random matrices}} (\bibinfo{publisher}{Academic Press},
  \bibinfo{address}{New York}, \bibinfo{year}{1991}).

\bibitem[{\citenamefont{Barnes}(1900)}]{Barnes00}
\bibinfo{author}{\bibfnamefont{E.~W.} \bibnamefont{Barnes}},
  \bibinfo{journal}{Q. J. Pure Appl. Math.} \textbf{\bibinfo{volume}{31}},
  \bibinfo{pages}{264} (\bibinfo{year}{1900}).

\bibitem[{\citenamefont{Forrester and Rains}()}]{ForresterRains02b}
\bibinfo{author}{\bibfnamefont{P.~J.} \bibnamefont{Forrester}}
  \bibnamefont{and} \bibinfo{author}{\bibfnamefont{E.~M.} \bibnamefont{Rains}},
  \eprint{math-ph/0211042}.

\bibitem[{\citenamefont{Porter and Stirling}(1990)}]{PorterStirling}
\bibinfo{author}{\bibfnamefont{D.}~\bibnamefont{Porter}} \bibnamefont{and}
  \bibinfo{author}{\bibfnamefont{D.}~\bibnamefont{Stirling}},
  \emph{\bibinfo{title}{Integral equations}} (\bibinfo{publisher}{Cambridge
  University Press}, \bibinfo{address}{Cambridge}, \bibinfo{year}{1990}).

\bibitem[{\citenamefont{Gradsteyn and Ryzhik}(1994)}]{Gradstein}
\bibinfo{author}{\bibfnamefont{I.~S.} \bibnamefont{Gradsteyn}}
  \bibnamefont{and} \bibinfo{author}{\bibfnamefont{I.~M.}
  \bibnamefont{Ryzhik}}, \emph{\bibinfo{title}{Table of Integrals, Series, and
  Products}} (\bibinfo{publisher}{Academic Press}, \bibinfo{address}{New York},
  \bibinfo{year}{1994}), \bibinfo{edition}{5th} ed.

\bibitem[{\citenamefont{Fisher and Hartwig}(1968)}]{FisherHartwig68}
\bibinfo{author}{\bibfnamefont{M.~E.} \bibnamefont{Fisher}} \bibnamefont{and}
  \bibinfo{author}{\bibfnamefont{R.~E.} \bibnamefont{Hartwig}},
  \bibinfo{journal}{Adv. Chem. Phys.} \textbf{\bibinfo{volume}{15}},
  \bibinfo{pages}{333} (\bibinfo{year}{1968}).

\bibitem[{\citenamefont{Bottcher and Silbermann}(1989)}]{BottcherSilbermann89}
\bibinfo{author}{\bibfnamefont{A.}~\bibnamefont{Bottcher}} \bibnamefont{and}
  \bibinfo{author}{\bibfnamefont{B.}~\bibnamefont{Silbermann}},
  \emph{\bibinfo{title}{Analysis of Toeplitz operators}}
  (\bibinfo{publisher}{Akademie}, \bibinfo{address}{Berlin},
  \bibinfo{year}{1989}).

\bibitem[{\citenamefont{Forrester and Pisani}(1992)}]{ForresterPisani}
\bibinfo{author}{\bibfnamefont{P.~J.} \bibnamefont{Forrester}}
  \bibnamefont{and} \bibinfo{author}{\bibfnamefont{C.}~\bibnamefont{Pisani}},
  \bibinfo{journal}{Nucl. Phys. B.} \textbf{\bibinfo{volume}{374}},
  \bibinfo{pages}{720} (\bibinfo{year}{1992}).

\bibitem[{\citenamefont{Hirschmann}(1965)}]{Hirschmann65}
\bibinfo{author}{\bibfnamefont{I.~I.} \bibnamefont{Hirschmann}},
  \bibinfo{journal}{Amer. J. of Math.} \textbf{\bibinfo{volume}{88}},
  \bibinfo{pages}{577} (\bibinfo{year}{1965}).

\bibitem[{\citenamefont{Johansson}(1997)}]{Johansson98}
\bibinfo{author}{\bibfnamefont{K.}~\bibnamefont{Johansson}},
  \bibinfo{journal}{Ann. of Math.} \textbf{\bibinfo{volume}{145}},
  \bibinfo{pages}{519} (\bibinfo{year}{1997}).

\bibitem[{\citenamefont{Beenakker}(1994)}]{Beenakker93}
\bibinfo{author}{\bibfnamefont{C.~W.~J.} \bibnamefont{Beenakker}},
  \bibinfo{journal}{Nucl. Phys. B.} \textbf{\bibinfo{volume}{422}},
  \bibinfo{pages}{515} (\bibinfo{year}{1994}).

\bibitem[{\citenamefont{Luke}(1969)}]{Luke}
\bibinfo{author}{\bibfnamefont{Y.~L.} \bibnamefont{Luke}},
  \emph{\bibinfo{title}{The special functions and their approximations}},
  vol.~\bibinfo{volume}{1} (\bibinfo{publisher}{Academic Press},
  \bibinfo{address}{New York}, \bibinfo{year}{1969}).

\end{thebibliography}

\end{document}